\documentstyle[preprint,prb,aps]{revtex}
\tighten
\begin{document}
\draft
\title{
Canonical Transformation Approach to the Ultrafast Non--linear 
Optical Dynamics of Semiconductors}
\author{I. E. Perakis and T. V. Shahbazyan}
\address{Department of Physics and Astronomy,  
Vanderbilt University Nashville, TN 37235}

\maketitle

\begin{abstract} 

We develop a  theory describing
the effects  of  many--particle Coulomb correlations
on the coherent ultrafast nonlinear optical response 
of semiconductors and metals.
Our approach is based on a mapping of
the nonlinear optical response of the ``bare'' 
system
onto the linear response of a ``dressed'' 
system.
The latter is 
characterized by effective time--dependent
optical transition matrix elements, 
electron/hole dispersions, 
and interaction potentials, 
which in undoped semiconductors 
are determined by 
the single--exciton and two--exciton 
Green functions   in the absence of optical fields. 
This mapping is achieved by eliminating the 
optically--induced charge fluctuations 
from the Hamiltonian using a Van Vleck canonical 
transformation. 
It takes into account  all 
many--body contributions up 
to a given order in the  optical 
fields  as well as 
important 
Coulomb--induced quantum dynamics 
to all orders in the optical 
field. 
Our approach  allows us to distinguish
between optical nonlinearities of different origins 
and provides a physically--intuitive 
interpretation of
their manifestations in ultrafast coherent nonlinear 
optical spectroscopy.
\end{abstract} 

\section{Introduction}  	

In semiconductors and metals, the Coulomb interactions
among  electrons  and 
holes strongly affect the optical spectra 
close to the onset of absorption.\cite{rev}  
In undoped semiconductors, 
the {\em e--h} attraction 
leads to exciton resonances,  while in 
doped quantum wells and metals 
it leads   to the Fermi-Edge-Singularity.\cite{exper}  
In the vicinity of  the  resonances, 
the  interactions 
between 
photoexcited  e--h pairs
and Fermi sea electrons  
also affect strongly
the time--dependence 
of the e--h  polarization 
measured in ultrafast nonlinear optics experiments.
This allows one to obtain new information on 
many--body correlations, not easily accessible 
with other experimental 
techniques,
by  using 
transient pump--probe and wave--mixing  spectroscopy,
\cite{rev1}   
The interpretation of such experiments, however, requires
solving  a challenging 
time-dependent many--body problem. 
Indeed, during femtosecond time scales, 
the Coulomb interactions 
dominate the time--dependence 
of the experimentally--measurable polarization.
In the coherent limit, the standard quasi--equilibrium 
approaches no longer apply.
This 
led to the development 
of alternative approximations
for  undoped semiconductors. The most important of those  
is the  Semiconductor Bloch
Equations (SBE's) approach, 
which treats 
exciton--exciton interaction effects 
on the nonlinear optical response  
within the Hartree--Fock approximation.\cite{sbe,koch}

Recent experiments, however, revealed  pronounced features 
in the amplitude and phase dynamics of the 
nonlinear polarizations that could not be  captured by the SBE's.\cite{muk}  
In Ref. 7, 
for example,
pronounced features in the phase and amplitude 
dynamics of the four--wave--mixing polarization 
were attributed 
to non--Markovian 
dephasing 
due to the Coulomb correlations,
neglected within the Hartree--Fock 
approximation.
In Ref. 8, 
it was shown that 
the  Boltzmann picture 
of relaxation 
is inadequate for 
extremely short 
optical pulse durations.
In Ref. 9, 
it was observed 
that the time decay and spectral width of the four--wave mixing 
polarization did not  correlate 
with the decay
of the time-integrated signal, 
as would have been expected based on the SBE's. 
Experimental evidence for 
the  effects of the bound biexciton state (not described by the SBE's) 
was presented, e.g., in 
Ref. 10. 
Correlation effects 
due to the repulsive 
exciton--exciton interaction 
were found to  significantly enhance the 
four--wave--mixing signal
for negative time delays
as function of magnetic field,
even in the absence of bound biexciton states.\cite{x--x}
Excitation--induced 
dephasing effects were observed, e.g., in 
Ref. 12. 

Experiments such as those mentioned above 
point out the need to incorporate in the theory  
exciton--exciton correlations
beyond the time--dependent Hartree--Fock approximation. 
This problem has been addressed 
within the  equations--of--motion approach  
for density matrices, which,
to a given order in the optical fields,
leads to a closed system of differential
equations that determine 
the  nonlinear polarizations.\cite{muk,eom,sham}
The solution  of such   equations,
however, must address the 
formidable  four--body problem 
of two electrons interacting with two holes.
With the exception  of one--dimensional systems \cite{sham}, 
the determination of such biexciton states is numerically 
very demanding. 
Various approximation schemes, devised mainly for calculations of the
third--order polarization,  $\chi^{(3)}$, 
have been recently proposed.\cite{muk,eom,sham}
The e--h Coulomb correlations also strongly affect the 
{\em coherent} polarization femtosecond dynamics
in doped quantum wells.\cite{acfes,ilias} Such effects originate
from the 
dynamical Fermi sea response to the 
photoexcited valence holes,\cite{rev,exper}
and can be observed for frequencies 
close to the Fermi enegy, 
where the absorption spectrum is dominated by the Fermi--edge
singularity
and the {\em e--e} quasiparticle scattering 
is inhibited.\cite{kim,hawr1,films}
The ultrafast dynamics of the Fermi--edge singularity 
was addressed
in Ref. 16. 

In this paper, we develop a new approach for describing correlation
effects in  ultrafast nonlinear spectroscopy. 
Our approach is motivated 
by the fact that 
the  nonlinear polarization
measured in pump--probe or four--wave--mixing experiments
arises from the linear  response to the probe optical field 
(propagating along direction ${\bf k_{\tau}}$)
of a system described by the Hamiltonian $H + H_{p}(t)$, 
where $H$ is the two--band interacting  Hamiltonian 
in the absence of optical fields 
and $H_{p}(t)$ describes the coupling to the pump optical field
(propagating along direction ${\bf k_{p}}$).
Terms nonlinear  in the probe optical field 
do not contribute to 
the third--order polarizations measured along directions 
${\bf k_{\tau}}$ (pump--probe) 
or $2 {\bf k_{p}} - {\bf k_{\tau}}$ (wave--mixing). They 
can also be neglected when the probe is  weaker than the pump 
field, as is often the case experimentally.
However, unlike the ``bare'' Hamiltonians $H$,
the Hamiltonian  $H + H_{p}(t)$
does not conserve the number of {\em e--h} pairs. 
The idea is to  ``block--diagonalize''
the Hamiltonian $H + H_{p}(t)$ in the basis of states with fixed number 
of {\em e--h} excitations, such as,  e.g., exciton states, biexciton states
etc. This can be achieved to any given
order in the optical fields by using Van Vleck 
canonical  transformations,\cite{van,cohen} generalized to the
time--dependent case. This way we obtain an effective 
{\em time--dependent} Hamiltonian, $H_{\rm eff}(t)$ which,
{\em does} conserve
the number of {\em e--h} excitations. Note that 
for the time--independent case, such 
a procedure is analogous to the  
Schrieffer--Wolff transformation used to  eliminate the 
charge fluctuations in the Anderson Hamiltonian,\cite{ander,mahbook}
where $H_{\rm{eff}}$ corresponds to the 
Kondo Hamiltonian.

The above procedure allows us to 
cast the experimentally--measurable 
nonlinear  polarizations 
in the  form of the  linear polarization of an 
effective two--band  system. The latter  is 
characterized by time--dependent band dispersion relations,  
interaction potentials, and transition matrix elements, which
can be obtained to 
any given order in the pump optical field.
In particular, as we will
show, to  second order in the pump field the above effective
parameters 
can be expressed in terms of the one-- and two--exciton Green functions 
for the ``bare''  Hamiltonian $H$.  Such a formulation 
allows us to interpret the various  dynamical 
features in the nonlinear absorption spectra, 
originating from the correlation effects, 
within the familiar framework 
developed for linear spectroscopy.

In particular, in linear absorption, 
the strength and spectral position of 
the excitonic resonances are determined by 
(i)
the dipole transition matrix elements
and (ii) the time evolution of the 
{\em e--h} pair  wavefunction. 
Here we 
establish a similar picture 
for the nonlinear absorption spectrum, which 
applies for any  pulse duration:
(i) 
The {\em e--h} pair photoexcited by the probe 
interacts with 
the {\em e--h} pairs photoexcited 
by the pump via  the Pauli principle
(Phase Space Filling effect), static 
exciton--exciton interactions
(Hartree--Fock contribution),  
 and exciton--exciton scattering processes that involve the exchange
of center--of--mass momentum (correlation effects). 
Such effects are described by a  renormalized transition  matrix
element. 
(ii) Similar
 interactions also lead 
to a pump--induced renormalization of the
semiconductor band 
energies and {\em e--h} interaction potential,
which 
affect the 
time evolution of the
(probe--induced) 
{\em e--h} pair
wavefunction. The corresponding 
time evolution operator that 
creates  the excitonic resonance 
 is governed by  
the Hermitean  effective 
Hamiltonian $H_{\rm eff}(t)$  which, importantly,  has  the same form
as $H$.
Therefore, the nonlinear absorption spectrum 
can  be viewed as arising from the 
probe--induced optical transitions to 
renormalized (``dressed'') exciton  states, which are 
determined by the  Schr\"{o}dinger equation 
with  Hamiltonian $H_{\rm eff}(t)$.
One, thus, should distinguish between  
two physically--distinct origins of the optical nonlinearity: 
optically--induced renormalizations of the 
transition matrix elements
and time--dependent parameter renormalizations in the 
Hamiltonian.
The latter  not only 
shift the  bound state energy, as would have been the case 
within the two--level system approximation,  but also changes its 
wavefunction and Bohr radius that determine the exciton oscillator
strength.  
Importantly,
 due to the  Coulomb interactions,
 $H_{\rm eff}(t)$ 
does not commute with 
itself at different times
(unlike for monochromatic photoexcitation),  which 
leads to quantum dynamics 
and memory effects
that, as we show,  are not captured by 
the third--order nonlinear response ( $\chi^{(3)}$)
or by using few--level system truncations. 
Here we provide a systematic method for including  all the above 
effects in a physically--intuitive fashion.

The paper is organized as follows. 
In Section II we set up the problem and briefly outline our final 
results. In Section III we describe the  canonical trasformations that 
eliminate all optically--induced charge fluctuations up to the second
order in the pump field. In Section IV we obtain the linear response 
of the system to the probe optical field.
In Section V we  focus on the pump--probe nonlinear  polarization 
and identify  the 
effective Hamiltonian
and optical transition operator. 
In Section VI
we  derive the four--wave--mixing polarization.
Section VII concludes the paper.

\section{\bf Problem Set--Up}

In the absence of optical fields, 
the excitation spectrum 
of a two-band semiconductor 
is described by the 
Hamiltonian 

\begin{equation} 
H =  \sum_{\bf {\bf q}} \varepsilon_{{\bf {\bf q}}}^{v} 
 b_{-{\bf {\bf q}}}^{\dag}\, b_{-{\bf {\bf q}}} \  + \
\sum_{{\bf {\bf q}}} (\varepsilon_{{\bf {\bf q}}}^{c}+E_g) \ a^{\dag}_{{\bf {\bf q}}}
 \,  a_{{\bf {\bf q}}}
 +V_{ee} +V_{hh}+V_{eh},
 \label{H0}
\end{equation}
where  $a^{\dag}_{{\bf {\bf q}}}$ is the creation operator
of a conduction electron
with energy $\varepsilon_{{\bf {\bf q}}}^{c}$
and mass $m_{e}$,
$b^{\dag}_{-{\bf {\bf q}}}$ is the creation operator
of a valence hole
with energy $\varepsilon_{{\bf {\bf q}}}^{v}$
and mass $m_{h}$,  
$V_{ee}, V_{eh}$, and $V_{hh}$
describe the
{\em e--e}, {\em e--h}, and {\em h--h} interactions,
respectively, and $E_g$ is the bandgap.
For simplicity, we consider a single parabolic 
valence band and suppress the spin degrees of freedom.

Ultrafast pump--probe and wave--mixing spectroscopy measures
the response of such a  system 
to photoexcitation by a pump optical field,
$E_{p}(t) e^{i {\bf k_{p}} \cdot {\bf r}  - i\omega_{p} t}$ 
(with $E_{p}(t)$ centered at $t=0$),
and a probe optical field, $E_{\tau}(t) 
e^{i{\bf k_{\tau}} \cdot {\bf r} - i\omega_{p} (t-\tau)}$
(with $E_{\tau}(t)$ centered at $t=\tau$), 
propagating along  directions
${\bf k_{p}}$ and ${\bf k_{\tau}}$, respectively,
and have central frequency $\omega_{p}$ and 
time--delay $\tau$.
In the rotating wave approximation,\cite{eber} 
the coupling to the optical field 
is described  by the Hamiltonian $H_{p}(t)$ (pump),

\begin{equation}
H_{p}(t) = d_{p}(t) \left[ e^{i {\bf k_{p}} \cdot
 {\bf r }}U^{\dag}+ H. c. \right], 
 \end{equation} 
and the Hamiltonian $H_{\tau}(t)$ (probe), 

\begin{equation}
H_{\tau}(t) = d_{\tau}(t) \left[ e^{i {\bf k_{\tau}} \cdot
 {\bf r } + i \omega_{p} \tau}U^{\dag}+ H. c. \right],
\end{equation} 
where
$U^{\dag} =
\sum_{{\bf {\bf q}}} a^{\dag}_{{\bf {\bf q}}} b^{\dag}_{-{\bf {\bf q}}}$ 
is the ``bare'' optical transition
operator, $d_{i}(t) = \mu E_{i}(t), i=p, \tau$ 
are the dipole energies, and  
$\mu$ is the interband transition 
matrix element.
The total Hamiltonian, including the
coupling to the optical fields, is 

\begin{equation} 
H_{\rm{tot}}(t) = H + H_{p}(t)+ H_{\tau }(t)
-\hbar \omega_{p}  
\sum_q
a_{{\bf {\bf q}}}^{\dag}\, a_{{\bf {\bf q}}}.
\end{equation} 
In the following, we will absorb the last term into $H$.

The pump--probe and four--wave--mixing signals are determined by 
the polarization

\begin{equation}\label{polar}
P(t) = \mu e^{-i\omega_{p}t}\langle \Psi(t) | U | \Psi(t) \rangle, 
\end{equation} 
where the state $|\Psi(t)\rangle$ satisfies the time-dependent
Schr\"{o}dinger equation with the Hamiltonian $H_{\rm{tot}}(t)$.
In the following, we distinguish between 
pump and probe optical fields 
by concentrating 
on the   polarizations that propagate along direction ${\bf
k}_{\tau}$
(pump--probe) and  
$2 {\bf k}_{p} - {\bf k}_{\tau}$ (four--wave--mixing). 
For a weak probe, the nonlinear 
polarizations arise from 
the linear response to the 
probe--induced  perturbation $H_{\tau}(t)$ 
of the coupled--two--band system 
system described by the time--dependent Hamiltonian 
$H + H_{p}(t)$. 
Note that, for $\chi^{(3)}$ 
along the above dierctions, 
the above is also true 
even if the probe and pump fields have comparable 
strength. 
In contrast to $H$,
this  Hamiltonian does not conserve the number of
{\em e--h} pairs, so we  transform it into an effective 
Hamiltonian $H_{\rm{eff}}(t)$ that has this property by  using 
canonical (Van--Vleck) transformations.\cite{van,cohen,ander}

Specifically, the Hilbert space of 
the bare semiconductor states (in the absence of optical fields) 
consists of disconnected subspaces,
which  can be labelled by the number of {\em e--h} airs. 
In this basis, $H$  has   block--diagonal form. 
The Hamiltonian $H_{p}(t)$ 
couples the different 
subspaces by creating/annihilating e--h pairs.
A unitary transformation that ``block--diagonalizes''
the Hamiltonian $H+ H_{p}(t)$, e.g., up to the second order in pump
field, has the form
$e^{-S_{2}}e^{-S_{1}}
[H+ H_{p}(t)]e^{S_{1}}e^{S_{2}}$, where the antihermitian
operators 
$S_{1}(t)$ and $S_{2}(t)$
create/annihilate 
one and two e--h pairs, respectively.
In order to achieve this in higher orders 
in the pump field, 
we should also include in $S_{1}(t)$ and $S_{2}(t)$
processes that create/annihilate 
odd and even numbers of e--h pairs, respectively. 
Correspondingly, 
the state $|\Psi(t) \rangle$ is given by 
 
\begin{equation}
|\Psi(t) \rangle = 
e^{S_{1}(t) }
e^{S_{2}(t)} 
|\Phi(t) \rangle \label{wav}
\end{equation}
where all 
the effects of the probe optical field are contained  
in the  state $|\Phi(t) \rangle$.
After the transformation operators $S_{i}^{(n)}(t)$ are calculated
and the effective Hamiltonian $H_{\rm{eff}}(t)$ is derived, the
polarization Eq.\ (\ref{polar}) with $| \Psi(t) \rangle$ given by 
Eq.\ (\ref{wav}) can be found in the first order in the probe field
by identifying all contributions 
that propagate along directions 
${\bf k_{\tau}}$ (pump--probe) or $2 {\bf k}_{p} - {\bf k}_{\tau}$
(four--wave--mixing).

Let us summarize  our  results
for undoped semiconductors. 
The pump-probe nonlinear
polarization, $P_{{\bf k}_{\tau}}$ can be presented as a sum of two terms,
$P_{{\bf k}_{\tau}}(t) = P^{(\rm{exc})}_{{\bf k}_{\tau}}(t) + 
P^{(\rm{biexc})}_{{\bf k}_{\tau}}(t)$,
where 
$P_{{\bf k}_{\tau}}^{(\rm{exc})}(t)$ 
and $P_{{\bf k}_{\tau}}^{(\rm{biexc})}(t)$ are polarizations  
due to the transitions from the full valence band ground state 
to  exciton and biexciton states, 
respectively. 
In particular,

\begin{equation} 
 P_{{\bf k}_{\tau}}^{(\rm{exc})}(t) =
 - \frac{i}{\hbar}   
e^{i {\bf k_{\tau}} \cdot {\bf r}- i\omega_{p} (t - \tau)}
\int_{-\infty}^{t} dt' d_{\tau}(t')
 \langle 0|
U_{\rm{exc}}(t) {\cal U}(t,t') U^{\dag}_{\rm{exc}}(t')
|0  \rangle, \label{spec1}
\end{equation}
where

\begin{equation} 
U^{\dag}_{\rm{exc}}(t)|0 \rangle  = \sum_{{\bf k}} M_{{\bf k}}(t) 
a^{\dag}_{{\bf k}} b^{\dag}_{-{\bf k}} |0 \rangle 
\end{equation} 
is the effective optical transition operator
to exciton states, 
$M_{{\bf k}}(t)$ being the corresponding matrix element. Here 
${\cal U}(t,t')$ is the 
time--evolution operator
satisfying  the Schr\"{o}dinger equation

\begin{equation}
i \hbar \frac{ \partial}{\partial t}
{\cal U} (t,t') 
= H_{\rm eff}(t)
{\cal U} (t,t') 
\label{eigen},
\end{equation}
where $H_{\rm{eff}}(t)$ is
the  effective semiconductor  Hamiltonian that acts on the
single {\em e--h } pair subspace. We show that $H_{\rm{eff}}(t)$
can be cast in  the same form 
as $H$:

\begin{equation}
H_{\rm{eff}}(t) = 
\sum_{{\bf {\bf q}}} \varepsilon_{{\bf {\bf q}}}^{c}(t) a_{{\bf {\bf q}}}^{\dag} a_{{\bf {\bf q}}}
+  \sum_{{\bf {\bf q}}} \varepsilon_{{\bf {\bf q}}}^{v}(t) b_{-{\bf {\bf q}}}^{\dag} 
b_{-{\bf {\bf q}}} 
  -\sum_{{\bf k}k'q } \,
\upsilon_{eh}({\bf k},{\bf k}';t)
a^{\dag}_{{\bf k}} \,
 b^{\dag}_{-{\bf k}}  \,  b_{-{\bf k}'} \, 
a_{{\bf k}'}. \label{Heff}
\end{equation}
Explicit expressions 
for the renormalized matrix elements and
parameters of $H_{\rm{eff}}(t)$ 
in terms of the single-- and two--exciton correlation
functions are derived in Section V.
The biexcitonic contribution to the pump--probe 
nonlinear polarization 
is given by an expression
similar to 
Eq.\ (\ref{spec1}), with the corresponding two--pair transition operator
$U^{\dag}_{\rm{biexc}}(t)$ (see Section V).

Note that in the absence of pump--induced photoexcitations,  $E_{p}(t) =0$,
the familiar form of the linear
absorption spectrum\cite{mahbook} is simply 
recovered 
from Eq.\ (\ref{spec1}) by setting 
 ${\cal U}(t,t') \rightarrow
e^{-\frac{i}{\hbar}  H (t - t')}$ and 
$U^{\dag}_{\rm{exc}}(t) \rightarrow U^{\dag}$.
The similar second--quantized forms of
Hamiltonians 
 $H_{\rm{eff}}(t)$ and $H$
suggest an interpretation  of the excitonic effects in the
nonlinear absorption spectrum in terms of 
probe--induced optical transitions 
to ``dressed'' exciton states. 
This analogy becomes more clear 
if one expresses the time-evolution operator ${\cal U}(t,t')$
in the basis of states
$|\Phi_{n}^{(1)}(t) \rangle$,

\begin{equation}
{\cal U}(t,t') =
\sum_{n} |\Phi_{n}^{(1)}(t) \rangle
  \langle 
\Phi_{n}^{(1)}(t')|,\label{prop}
\end{equation}
which 
satisfy  the Schr\"{o}dinger equation 
for the Hermitean Hamiltonian $H_{\rm eff}(t)$ 
with the initial condition that, before the pump pulse, 
$|\Phi_{n}^{(1)}(t) \rangle$ coincide with 
the exciton eigenstates of $H$.

As we will show in the next Section, the time--dependence  of
the parameters in the  
effective Hamiltonian Eq.\ (\ref{Heff})
follows that of  the pump amplitude. Therefore, as 
can be seen from  Eq.\ (\ref{eigen}), one can replace 
${\cal U} (t,t')$ by  $e^{- i H (t-t')/\hbar}$ 
when the probe comes at time $t'=\tau$ 
{\em after} the pump.
Thus the optically--induced corrections in 
$H_{\rm{eff}}(t)$ 
affect the pump--probe spectra only for negative 
probe delays or when the two pulses overlap in time. 
Furthermore, their effects 
are  most  pronounced 
in pump--probe spectroscopy,
where, unlike in four--wave--mixing, 
they already contribute to  the third--order 
nonlinear polarization.
On the other hand, the optically--induced renormalization in the
matrix element $M_{{\bf k}}(t)$, 
which includes contributions from  the Phase Space Filling 
as well as  from  the exciton--exciton 
interactions (see Section V), 
is finite even after the  
pump optical field is gone  
and therefore dominates the pump--probe 
signal for positive probe delays. 
The same is true for the biexcitonic contribution
to the pump--probe nonlinear polarization, 
which, as discussed below, vanishes for negative time delays. 

\section{\bf Photoexcitation by the Pump Optical Field}
\subsection{Elimination of single--pair transitions}

Our goal is to ``block-diagonalize'' the Hamiltonian $H+H_{p}(t)$
by performing the canonical
transformation Eq.\ (\ref{wav}).
In this subsection we determine 
the transformation operator 
$S_{1}(t) $ that eliminates the single-pair pump-induced
transitions. We start with a 
slightly more general problem,
useful for calculating higher order nonlinear
polarizations, 
by deriving a differential equation for the  operator $S(t)$ 
which ensures that all {\em odd}--pair charge fluctuations 
are eliminated from the transformed Hamiltonian.\cite{van}

In the time--dependent 
Schr\"{o}dinger equation 
$[i\hbar\partial/\partial t -H_{tot}(t)]|\Psi(t) \rangle=0$ we substitute 
$|\Psi(t) \rangle=e^{S(t)}|\chi(t) \rangle$
and, after  acting with the  operator $e^{-S(t) }$ 
on its lhs, we get 

\begin{equation} 
e^{-S(t) }
\left[ i \hbar 
 \frac{\partial}{\partial t}
- H \right] e^{S(t) } | \chi(t) \rangle 
=  e^{-S(t) } \left[ H_{p}(t)  + H_{\tau}(t) \right] 
 e^{S(t) } | \chi(t) \rangle
\label{schroed}. 
\end{equation} 
Since both $S(t)$ 
and $H_{p}(t)$ 
create/annihilate 
odd numbers of  {\em e--h} pairs
and $H$ conserves the number of {\em e--h} pairs, 
all odd--pair contributions 
to $|\chi(t) \rangle $  
originate from terms 
in the Taylor expansion of
the exponentials in
Eq.\ (\ref{schroed}) 
of the order of 
$\left(S\right)^{2n+1} 
\left(i \hbar 
 \frac{\partial}{\partial t}
- H \right)$ 
or  $\left(S \right)^{2n} H_{p}$, 
$n$ being an integer.
Following Ref. 22, 
we determine $S(t)$ by requiring that 
all such terms, which describe pump--induced charge fluctuations, 
cancel out. 

Using 
the Baker--Campbell--Hausdorff expansion for an arbitrary
operator $A(t)$,

\begin{equation}
 e^{-S(t)} A(t) 
 e^{S(t) }  = 
\left[ \cosh \hat{S}(t)
 + \sinh \hat{S}(t) 
\right] A(t), \label{baker} 
\end{equation}
where 
$\hat{S}(t)$ is
a superoperator  defined by 
$\hat{S}(t) A(t) = [A(t), S(t)]$, we find that 
$S(t)$ should satisfy

\begin{equation}
i \hbar \sinh[\hat{S}(t)] 
\frac{\partial}{\partial t} = 
\sinh[\hat{S}(t)] H + 
\cosh[\hat{S}(t)] H_{p}(t). \label{ap1}
\end{equation}
Such a form is inconvenient, however, since 
the lhs includes 
multiple commutators  
of $S(t)$
with its time-derivatives.
These can be eliminated by applying  
$\hat{S}(t) 
\sinh^{-1} \hat{S}(t)$
(defined by its Taylor expansion) 
to  both sides of Eq.\ (\ref{ap1}).
Using the relation 
$\hat{S} 
\frac{\partial}{\partial t} =
\frac{\partial S}{\partial t}$, 
we  obtain 

\begin{equation}
i \hbar \frac{\partial 
S(t) 
}{\partial t} = 
\left[  H, S(t) \right]  + \hat{S}(t)  
\coth [ \hat{S}(t)]  H_{p}(t), \label{van1}
\end{equation}
with the initial condition 
$S(- \infty) = 0$. 

Eliminating 
all contributions to  Eq.\ (\ref{schroed})
with odd numbers of {\em e--h} pairs by 
using Eq.\ (\ref{van1}),
we obtain after collecting the 
remaining terms

\begin{equation}
 \cosh[\hat{S}(t)]
\left[ i \hbar\frac{\partial}{\partial t}
- H \right]  | \chi(t) \rangle = 
 \sinh[\hat{S}(t)] \ H_{p}(t) \
| \chi(t) \rangle 
+ e^{-S(t) } H_{\tau}(t) 
 e^{S(t) } | \chi(t) \rangle.
\end{equation} 
This equation, however, is not particularly useful 
because it does not have the form of a 
Schr\"{o}dinger equation.
After some manipulations involving  adding and subtracting 
the operator 
$i \hbar \frac{\partial}{\partial t} - H$ 
and inserting the identity 
$\left(\hat{S}\right)^{-1} 
\hat{S} = 1$, 
it can be rewritten in the form 

\begin{eqnarray}
\left( i \hbar 
\frac{\partial}{\partial t} 
-H \right) | \chi(t) \rangle
= - \left[ \cosh[\hat{S}(t)] - 1 \right]
\left[\hat{S}(t)\right]^{-1} 
\hat{S}(t)
\left( i \hbar \frac{\partial}{\partial t} - H
\right)  | \chi(t)
\rangle\nonumber\\
+ \sinh[\hat{S}(t)] H_{p}(t) | \chi(t) \rangle 
+  e^{-S(t) } H_{\tau}(t) 
 e^{S(t) } | \chi(t) \rangle 
\label{chi}.
\end{eqnarray}
We now simplify the first term on the  rhs of Eq.\ (\ref{chi}) 
by making use of  Eq.\ (\ref{van1}).
After some algebra, we finally obtain

\begin{equation}
i \hbar \frac{\partial 
|\chi(t) \rangle  }{\partial t} = 
\left[ H
+ \tanh \left[\frac{\hat{S}(t)}{2}\right] H_{p}(t) 
\right] \ |\chi(t) \rangle
+  e^{-S(t) } H_{\tau}(t) 
 e^{S(t) } | \chi(t) \rangle.
\label{van2} 
\end{equation}
Eq.\ (\ref{van2}) now has the form of the Schr\"{o}dinger equation 
for $| \chi(t) \rangle$,
with a 
time--dependent  Hamiltonian depending on $S(t)$. 
The above derivation generalizes the results of 
Ref.\ 22 
to  time--dependent systems. 

The operator $S_{1}$, which eliminates the single-pair excitations,
can be obtained by linearizing Eq.\ (\ref{van1}). 
The nonlinear in $S_{1}$ terms in Eqs.\ (\ref{van1})
and (\ref{van2})
describe  the corrections in the effective 
semiconductor 
parameters
of the order of $E_{p}^{4}$ or higher. 
Using  the decomposition of the anti--hermitian operator $S_{1}(t)$,

\begin{equation} 
S_{1}(t) = \Sigma_{1}(t) 
e^{-i {\bf k_{p}} \cdot {\bf r}} -
\Sigma_{1}^{\dag}(t) e^{i {\bf k_{p}} \cdot {\bf r}},
\label{unit} 
\end{equation}
where $\Sigma_{1}^{\dag}(t)$ and $\Sigma_{1}(t)$ create and annihilate
single {\em e--h} pairs, respectively, we obtain 
the following equation for 
$\Sigma_{1}^{\dag}(t)$: 

\begin{equation}
i \hbar \frac{\partial 
\Sigma_{1}^{\dag}(t)}{\partial t} = 
\left[ H,\Sigma_{1}^{\dag}(t) 
\right] - d_{p}(t) U^{\dag},
\label{sig}
\end{equation}
which has the formal solution
 
\begin{equation} 
\Sigma_{1}^{\dag}(t)
= \frac{i}{\hbar} 
\int_{-\infty}^{t} dt'
d_{p}(t')
e^{ i H (t'-t)/\hbar} U^{\dag} e^{-i H (t'-t)/\hbar}. 
\label{sigop}
\end{equation}
Note that, since the Hamiltonian $H$ 
conserves the number of {\em e--h} pairs
and the optical transition operator 
$U^{\dag}$ creates a single {\em e--h} pair, 
 $\Sigma_{1}^{\dag}(t)$ also 
creates a single {\em e--h} pair. 
Furthermore, since both
$H$ and $U^{\dag}$ conserve momentum, so does 
$\Sigma_{1}^{\dag}(t)$.

It is convenient to represent $\Sigma_{1}^{\dag}(t)$ as a sum of
one--body  and   many--body parts,
 
\begin{equation}\label{decomp}
\Sigma_{1}^{\dag}(t)= \Sigma_{\rm{\rm{x}}}^{\dag}(t) + 
\Sigma_{\rm{xx}}^{\dag}(t),
\end{equation}
where 

\begin{equation} 
 \Sigma_{\rm{\rm{x}}}^{\dag}(t) = 
\sum_{{\bf q}} P_{\rm{x}}({\bf q},t)
a^{\dag}_{{\bf q}} \ b^{\dag}_{-{\bf q}} 
\label{dir}
\end{equation} 
describes the photoexcitation of a single  {\em e--h} pair 
(or exciton) 
with zero total momentum, while 
the many--body operator
$\Sigma_{\rm{xx}}^{\dag}(t)$ 
contains quartic and higher order terms
in the single--particle operators.
As we discuss later, in undoped semiconductors, 
$\Sigma_{\rm{xx}}^{\dag}(t)$ 
describes the exciton--exciton interactions. 
After substituting Eq.\ (\ref{decomp}) 
into Eq.\ (\ref{sig}) and collecting all
terms quadratic 
in the single particle operators, we obtain

\begin{equation} 
i \hbar \frac{\partial}{\partial t} \
P_{\rm{x}}({\bf q},t) 
= \left( \Delta \Omega + \varepsilon_{{\bf q}}^{c}
+ \varepsilon_{{\bf q}}^{v} \right) P_{\rm{x}}({\bf q},t) 
- \sum_{{\bf q}'} \upsilon({\bf q}-{\bf q}') P_{\rm{x}}({\bf q}',t)- d_{p}(t), 
\label{dir1} 
\end{equation}
where $\upsilon({\bf q})$ is the Coulomb potential and 
$\Delta \Omega=E_{g} - \hbar \omega_{p}$ is the detuning.
Thus, the amplitude $P_{\rm{x}}({\bf q},t)$ is just 
the linear polarization due to the pump field. 
This can also be seen by substituting 
Eq.\ (\ref{wav}) into the polarization 
equation  Eq.\ (\ref{polar}) (with $S_p^{(2)}=0$)
in the absence of the probe field.

Note that, due to the  interaction terms in $H$, the commutator
in the rhs of Eq.\ (\ref{sig}) generates 
the following terms, quartic in the single-particle
operators:

\begin{equation} 
 U^{\dag}_{\rm{xx}}(t)=
\sum_{{\bf k}{\bf q}} \upsilon({\bf q}) 
\left[ P_{\rm{x}}({\bf k},t)  - P_{\rm{x}}({\bf k}-{\bf q},t) 
\right] 
U^{\dag}_{\rm{xx}}({\bf k},{\bf q}),
\end{equation} 
with

\begin{equation} \label{Uxx}
U_{\rm{xx}}^{\dag}(k,{\bf q}) =
\sum_{{\bf k}'} \left(
a^{\dag}_{{\bf k}-{\bf q}}
b^{\dag}_{-{\bf k}} 
a^{\dag}_{{\bf k}'+{\bf q}} a_{{\bf k}'}- 
a^{\dag}_{{\bf k}-{\bf q}}
b^{\dag}_{-{\bf k}} 
b^{\dag}_{-{\bf k}'+  {\bf q}} b_{-{\bf k}'}\right).
\end{equation}
Thus, we get the following equation for 
 $\Sigma_{\rm{xx}}^{\dag}(t)$:

\begin{equation}
i \hbar \frac{\partial 
\Sigma^{\dag}_{\rm{xx}}(t)}{\partial t} = 
\left[ H,\Sigma^{\dag}_{\rm{xx}}(t) 
\right] + U^{\dag}_{\rm{xx}}(t),
\label{sig3}
\end{equation}
which has the formal solution 

\begin{equation}\label{sigin}
\Sigma_{\rm{xx}}^{\dag}(t)
= \frac{i}{\hbar} 
\sum_{{\bf k}{\bf q}} \upsilon({\bf q}) 
\int_{-\infty}^{t} dt'
\left[ P_{\rm{x}}({\bf k}-{\bf q},t')  - P_{\rm{x}}({\bf k},t') \right] 
U_{\rm{xx}}^{\dag}({\bf k},{\bf q},t-t'), 
\end{equation}
with

\begin{equation} 
U_{\rm{xx}}^{\dag}({\bf k},{\bf q},t) 
=e^{-i H  t/\hbar}
U_{\rm{xx}}^{\dag}({\bf k},{\bf q}) 
e^{i H t/\hbar}.
\end{equation} 
As can be seen from Eq.\ (\ref{sigin}),
$\Sigma_{\rm{xx}}^{\dag}(t)$ describes the photoexcitation 
of a single {\em e--h} pair accompanied by Coulomb interactions 
with additional photoexcited or Fermi sea carriers. 
Note that the time evolution 
of $\Sigma_{\rm{xx}}^{\dag}(t)$ 
is driven by the linear pump--induced {\em polarization}
and thus, in contrast to  $\Sigma_{\rm{x}}^{\dag}(t)$, only
indirectly depends  on the pump amplitude $d_{p}(t)$.  

We now turn  to the Schr\"{o}dinger equation 
Eq.\ (\ref{van2}).
The leading--order 
 corrections 
to the effective Hamiltonian 
are given 
by the terms linear in  $S_1$.
With the help of  Eq.\ (\ref{unit})
we obtain 

\begin{eqnarray}
\left[i \hbar \frac{\partial}{\partial t} 
-H_{\rm{eff}}(t)\right]|\chi(t) \rangle  = 
- \frac{d_{p}(t)}{2}
\left(e^{2 i {\bf k_{p}} {\bf r}} 
 \left[U^{\dag}, \Sigma_{1}^{\dag}(t) \right] + H.c. \right)
| \chi(t) \rangle
\nonumber\\
+  e^{-S_{1}(t) } H_{\tau}(t) 
 e^{S_{1}(t) } | \chi(t) \rangle. 
\label{first}
\end{eqnarray} 
where 
\begin{equation} 
H_{\rm{eff}}(t) = H - \frac{d_{p}(t)}{2}
\left(\left[U, \Sigma_{1}^{\dag}(t) \right] + H.c. \right)
 \label{eff} 
\end{equation} 
is a time--dependent effective  Hamiltonian
that  conserves the number of {\em e--h} pairs and
$\Sigma_{1}^{\dag}(t)$ is 
given by Eq.\ (\ref{sig}) with initial condition
$\Sigma_{1}^{\dag}(-\infty)=0$.
The first term in the rhs
of Eq.\ (\ref{first})
describes the pump--induced 
two--pair creation/annihilation processes. These can be eliminated
by performing a second canonical tranformation, as described in the
next subsection.

Concluding this subsection, let us address the condition of validity
of our approach.  It is useful 
to write down a formal solution of Eq.\ (\ref{sig}) 
in the basis of the
N--hole  many--body eigenstates, $|\alpha N \rangle$ with
energies  $E_{\alpha N}$, 
of the Hamiltonian $H$.
Here, for example, $N$=0 gives the 
semiconductor ground state $| 0 \rangle$,  
$|\alpha 1\rangle$ denotes the exciton eigenstates,
$|\alpha 2\rangle$ denotes the biexciton eigenstates,  etc.
In this basis, the solution of Eq.\ (\ref{sig}) can be written as

\begin{equation}
\frac{  \langle \beta N+1
|\Sigma_{1}^{\dag}(t) |\alpha N \rangle}{
 \langle \beta N+1| U^{\dag} |\alpha N \rangle }
=\frac{i}{\hbar}
\int_{-\infty}^{t} d_{p}(t')
e^{\frac{i}{\hbar}  (t'-t) \left( \Delta \Omega
+ \Delta E_{\alpha \beta}^{N} \right)}
e^{-\Gamma(t-t')}dt',  \label{sig1}
\end{equation}
where we separated out the detuning $\Delta \Omega$ and denoted
$\Delta E_{\alpha \beta}^{N}
= E_{\beta N+1}- E_{\alpha N}$;
the width $\Gamma$
describes the effects
of dephasing
processes not included in 
$H$ (e.g., due to phonons).
It can be seen that 
for resonant excitations (small $\Delta\Omega$) the rhs  is of the
order of $d_{p} t_{p}/\hbar$. Thus, for short pulses, this
parameter justifies the expansion in terms of
optical fields. Note that, for off-resonant excitation, this expansion
is valid even for longer pulse durations.

\subsection{Elimination of two--pair transitions} 

We now perform the second canonical transformation in order to
eliminate the first term in the rhs of Eq.\ (\ref{first}), which 
describes biexcitonic transitions. We define 

\begin{equation} 
| \chi(t)\rangle =  
e^{S_{2}(t)} 
|\Phi(t) \rangle, 
\label{chi1}
\end{equation} 
where, in the absence of the probe field, 
$ |\Phi(t)\rangle$ has fixed number of {\em e--h} pairs. 
Again, using the anti--Hermicity of $S_{2}(t)$, we decompose it
as 

\begin{equation} 
S_{2}(t) = \Sigma_{2}(t) 
e^{-2 i {\bf k_{p}} \cdot {\bf r}} -
\Sigma^{\dag}_{2}(t) e^{ 2i {\bf k_{p}} \cdot {\bf r}},
\label{unit2} 
\end{equation}
where $\Sigma^{\dag}_{2}(t)$ and $\Sigma_{2}(t)$ create and
annihilate two {\em e--h} pairs, respectively.
Substituting Eq.\ (\ref{chi1}) into Eq.\ (\ref{first}) and requiring
that all two--pair terms  cancel out, we obtain 
similar to the previous subsection 
the following
equation for $\Sigma^{\dag}_{2}(t)$ to the leading order, 

\begin{equation}
i \hbar \frac{\partial 
\Sigma^{\dag}_{2}(t)}{\partial t} = 
 \left[ H,\Sigma^{\dag}_{2}(t) 
\right] - \frac{d_{p}(t)}{2} [\Sigma_{1}^{\dag}(t),U^{\dag}],
\label{sig2}
\end{equation}
which has the solution
 
\begin{equation} 
\Sigma_{2}^{\dag}(t)
= \frac{i}{2\hbar} 
\int_{-\infty}^{t} dt'
d_{p}(t')
e^{ i H (t'-t)/\hbar} [\Sigma_{1}^{\dag}(t),U^{\dag}]e^{-i H (t'-t)/\hbar}. 
\label{sigop2}
\end{equation}
As we will show later, 
$\Sigma_{2}^{\dag}(t)$ 
includes all exciton--exciton interaction contributions to the 
third--order 
four--wave--mixing polarization 
but 
only affects the pump--probe signal 
via  higher order ($E_{p}^{4}$)
parameter renormalizations.

\section{ Probe--Induced Optical Transitions}

In the previous section, we ``block-diagonalized'' the Hamiltonian
$H+H_p(t)$ 
(up to the second
order in the pump field)
in the basis of states which, in the absence of the probe
optical field, conserve the number of {\em e--h} pairs.
  In the presence of the probe field, the state
 $| \Phi(t)\rangle$
(Eq.\ (\ref{chi1}))
 satisfies 
the Schr\"{o}dinger equation 

\begin{equation}
i \hbar \frac{\partial}{\partial t} 
| \Phi(t) \rangle = 
H_{\rm{eff}}(t)
| \Phi(t) \rangle
 + d_{\tau}(t) 
\left[ e^{i  {\bf k_{\tau}}{\bf r}+i\omega_{p}\tau} U^{\dag}_{\rm{r}}(t) 
+ H.c. \right] | \Phi(t) \rangle,
\label{sigprob}
\end{equation} 
where $H_{\rm eff}(t)$ is given by Eq.\ (\ref{eff}) and

\begin{equation} 
U^{\dag}_{\rm{r}}(t) =
e^{-S_{2}(t) }  e^{-S_{1}(t) } 
U^{\dag}   e^{S_{1}(t) } 
 e^{S_{2}(t) }  \label{Ur}
\end{equation} 
is the (transformed) optical transition operator.
Following the standard procedure, 
we obtain in the first  order in the probe field

\begin{equation} 
|\Phi(t) \rangle = 
\left\{ 1 - \frac{i}{\hbar} 
\int_{-\infty}^{t} 
d_{\tau}(t') 
{\cal U}(t,t')
\left[ e^{i  {\bf k_{\tau}} {\bf r}+i\omega_{p}\tau} U^{\dag}_{\rm{r}}(t') 
+ H.c. \right]
\right\} |\Phi^{(0)}(t') \rangle, \label{phi}
\end{equation} 
where ${\cal U}(t,t')$ is the time-evolution operator satisfying

\begin{equation}
i \hbar \frac{\partial}{\partial t}{\cal U}(t,t')=
H_{\rm{eff}}(t){\cal U}(t,t'), \label{evol} 
\end{equation}
and $|\Phi^{(0)}(t)\rangle = 
{\cal U}(t,-\infty) | 0 \rangle $ is the time--evolved 
ground state $|0 \rangle$.
Since $H_{\rm{eff}}(t)$ conserves the number of {\em e--h} pairs, 
$|\Phi^{(0)}(t)\rangle$ contains no {\em e--h} pairs. 
In undoped semiconductors, 
the  only such state is the ground state
so that
$ |\Phi^{(0)}(t)\rangle  \propto | 0 \rangle$.

Using Eq.\ (\ref{phi})  for $ |\Phi(t)\rangle$, the 
polarization 
$P(t) = \mu e^{-i\omega_{p}t}\langle \Psi(t)| U | \Psi(t) \rangle 
= \mu e^{-i\omega_{p}t}\langle \Phi(t)|U_{\rm{r}}(t) | \Phi(t) \rangle$
can be written as 

\begin{eqnarray} 
P(t) =
&&
-\frac{i\mu}{\hbar} e^{-i\omega_{p}t}
\int_{-\infty}^{t} 
d_{\tau}(t')
\nonumber\\ &&\times
\{
\langle 0| U_{\rm{r}}(t){\cal U}(t,t')
[ e^{i{\bf k_{\tau}} {\bf r}+i\omega_{p}\tau}
U^{\dag}_{\rm{r}}(t')
+  e^{-i  {\bf k_{\tau}} {\bf r}-i\omega_{p}\tau} 
U_{\rm{r}}(t')] |0\rangle
\nonumber\\ &&
-\langle 0| 
\left[e^{i  {\bf k_{\tau}} {\bf r}+i\omega_{p}\tau} U^{\dag}_{\rm{r}}(t') 
+ e^{-i  {\bf k_{\tau}} {\bf r}-i\omega_{p}\tau} U_{\rm{r}}(t')  \right]
{\cal U}(t',t)U_{\rm{r}}(t)|0\rangle
\}. 
\label{pol}
\end{eqnarray}
Explicit expressions for 
$ U^{\dag}_{\rm{r}}(t)$
are derived in the next sections. 

\section{Pump-Probe Spectrum}
\subsection{Pump--probe polarization} 

In this section we derive the  closed form expression for the
nonlinear polarization propagating in the direction of the 
probe optical field.
 In order to extract the pump-probe polarization from 
Eq.\ (\ref{pol}),
one should retain only 
contributions that are proportional to $e^{i {\bf k_{\tau}} {\bf r}}$
and are independent of the pump direction ${\bf k}_{p}$.
To the leading order,
the transition operator $U^{\dag}_{\rm{r}}(t)$
can be obtained 
by expanding 
Eq.\ (\ref{Ur})
in terms of  $S_{1}$ and
$S_{2}$ (given by Eqs.\ (\ref{unit}) and (\ref{unit2}))
using the Baker--Campbell--Hausdorff formula, Eq.\ (\ref{baker}):

\begin{equation} 
U^{\dag}_{\rm{r}}(t) = 
 U^{\dag}_{\rm{exc}}(t) + 
e^{- i {\bf k}_{p}{\bf r}}  U^{\dag}_{0}(t) 
 + U^{\dag}_{-1}(t) e^{- 2 i {\bf k}_{p}{\bf r}} 
+U^{\dag}_{\rm{biexc}}(t) e^{i {\bf k}_{p}{\bf r}} + U^{\dag}_{3}(t) 
e^{2 i {\bf k}_{p}{\bf r}}.
\end{equation}
(Note that the higher--order corrections 
 do not contribute also to $\chi^{(3)}$
measured along the probe direction.) 
Here the operator 

\begin{equation} 
U^{\dag}_{\rm{exc}}(t) 
= U^{\dag} + \frac{1}{2} 
\left[ \Sigma_{1}(t), \left[U^{\dag}, \Sigma_{1}^{\dag}(t)\right]
\right] + \frac{1}{2} \left[ 
\Sigma_{1}^{\dag}(t), \left[ U^{\dag}, \Sigma_{1}(t)\right]
\right],\label{1}
\end{equation} 
creates a single {\em e--h} pair; the operator 

\begin{equation}
U^{\dag}_{0}(t) = 
\left[ U^{\dag}, \Sigma_{1}(t) \right],
\end{equation}
conserves the number of {\em e--h} pairs; the operator 
 
\begin{equation} 
U^{\dag}_{\rm{biexc}}(t) = 
\left[\Sigma_{1}^{\dag}(t), U^{\dag}\right],
\label{U2}
\end{equation} 
creates two {\em e--h} pairs; the operator

\begin{equation} 
U^{\dag}_{-1}(t) = \left[ U^{\dag}, \Sigma_{2}(t) \right] 
+ \frac{1}{2} \left[ \left[ U^{\dag},\Sigma_{1}(t) \right],
\Sigma_{1}(t) \right], 
\end{equation} 
annihilates  a single {\em e--h} pair; and  the operator
\begin{equation} 
U^{\dag}_{3}(t)= 
\left[\Sigma_{2}^{\dag}(t), U^{\dag} \right] 
 + \frac{1}{2} \left[ \left[ U^{\dag}, \Sigma_{1}^{\dag}(t)\right], 
\Sigma_{1}^{\dag}(t) \right],
\end{equation}  
creates three {\em e--h} pairs. Furthermore, it is easy to see that 

\begin{equation} 
U_{\rm{r}}(t) |0\rangle 
= [e^{i {\bf k}_{p}{\bf r}}  U_{0}(t) 
+ U_{-1}(t) e^{2 i {\bf k}_{p}{\bf r}}]|0\rangle, 
\label{UU1}
\end{equation} 
and

\begin{equation} 
U_{\rm{r}}^{\dag}(t) |0 \rangle 
 =\left[  U^{\dag}_{\rm{exc}}(t) +
e^{- i {\bf k}_{p}{\bf r}}  U^{\dag}_{0}(t) 
+U^{\dag}_{\rm{biexc}}(t) e^{i {\bf k}_{p}{\bf r}} + U^{\dag}_{3}(t) 
e^{2 i {\bf k}_{p}{\bf r}} \right] |0 \rangle,
\label{UU2}
\end{equation} 
while  all other terms annihilate the ground state $|0\rangle$. 
Substituting Eqs.\ (\ref{UU1}) and (\ref{UU2}) into  Eq.\ (\ref{pol})
and retaining only terms proportional to 
$e^{i {\bf k_{\tau}} {\bf r}}$, we obtain
$P_{{\bf k}_{\tau}}(t) =
P^{(\rm{exc})}_{{\bf k}_{\tau}}(t) + 
P^{(\rm{biexc})}_{{\bf k}_{\tau}}(t)$,
where 

\begin{equation} 
 P_{{\bf k}_{\tau}}^{(\rm{exc})}(t) =
 - \frac{i\mu}{\hbar}   
e^{i{\bf k_{\tau}} \cdot {\bf r}- i\omega_{p} (t - \tau)} 
\int_{-\infty}^{t} dt' d_{\tau}(t') 
 \langle 0|
U_{\rm{exc}}(t) {\cal U}(t,t') U^{\dag}_{\rm{exc}}(t')
|0  \rangle, \label{PP1}
\end{equation}
and

\begin{equation} 
 P_{{\bf k}_{\tau}}^{(\rm{biexc})}(t) = - \frac{i\mu}{\hbar}   
e^{i{\bf k_{\tau}} \cdot {\bf r}- i\omega_{p} (t - \tau)} 
\int_{-\infty}^{t} dt' d_{\tau}(t')
 \langle 0|U_{\rm{biexc}}(t) {\cal U}(t,t') U^{\dag}_{\rm{biexc}}(t')
|0  \rangle. 
\label{PP2}
\end{equation}
In the derivation, all  terms containing $U_0$
cancelled each other out.
Note that the above expressions apply to both undoped and doped
semiconductors. 

We have thus expressed the nonlinear polarizations,  
Eqs.\ (\ref{PP1})--(\ref{PP2}), in terms of the linear response
to the probe optical field
of a system described by a Hermitean time--dependent effective Hamiltonian,
Eq.\ (\ref{eff}), that does not include any charge fluctutations.
The first term, Eq.\ (\ref{PP1}), descibes the excitonic
(single--hole state) contribution
to the pump--probe polarization. Eq.\ (\ref{PP1})
contributes to the pump--probe spectrum at frequencies 
close to the exciton energy.
Since $U^{\dag}_{\rm{biexc}}(t)$
creates two e-h pairs, the second term, Eq.\ (\ref{PP2}),
describes a biexcitonic contribution,
not included in the SBE's. 
Note that $\Sigma_{1}^{\dag}(t)$, and 
hence 
$ U^{\dag}_{\rm{biexc}}(t)$
vanishes if the probe pulse arrives 
before the pump pulse. 
Therefore $P_{{\bf k}_{\tau}}^{(\rm{biexc})}(t)$ 
is unimportant for negative time delays, 
for which the time--dependence of
$H_{\rm{eff}}(t)$ contributes the most. 
Biexciton (two--hole) states   
also  affect  $P_{{\bf k}_{\tau}}^{(\rm{exc})}(t)$ 
by acting as intermediate states 
in the renormalization of the parameters
entering into the optical transition operator
$U_{\rm{exc}}(t)$ and the 
effective Hamiltonian 
$H_{\rm{eff}}(t)$, as described later. 

In the following  sections, we 
focus on undoped semiconductors. 
We show that 
$U_{\rm{exc}}(t)$ and 
$H_{\rm{eff}}(t)$
have the same operator form
as their ``bare'' counterparts $U$ and
$H$, respectively; they only differ in the parameters.
This fact allows us to view the
non--linear polarization as arising 
from the linear response 
to a probe--induced
  optical transition 
from the ground state 
to a nonstationary ``dressed'' exciton state
evolving in time with Hamiltonian 
$H_{\rm{eff}}(t)$.

\subsection{Optical transition matrix elements}

In this subsection we derive explicit expressions 
for the effective transition matrix elements. 
We start by showing that,
when acting on 
the ground  state $| 0 \rangle$,
as in  Eq.\ (\ref{PP1}),
the operator
$U^{\dag}_{\rm{exc}}(t)$
has the same operator  form  as 
$U^{\dag}$.
Substituting 
$\Sigma_{1}^{\dag}(t)
= \Sigma_{\rm{x}}^{\dag}(t) + \Sigma_{xx}^{\dag}(t)$ 
into Eq.\ (\ref{1}) and using the second--quantized 
expression
for  $\Sigma_{\rm{x}}^{\dag}(t)$, Eq.\ (\ref{dir}), 
we obtain after some algebra

\begin{eqnarray} 
U^{\dag}_{\rm{exc}}(t)| 0 \rangle  
=
 \sum_{{\bf k}} \left[ 1 - \left|P_{\rm{x}}({\bf k},t)\right|^{2} \right] 
a^{\dag}_{{\bf k}} b^{\dag}_{-{\bf k}} | 0 \rangle
- \frac{1}{2} \left[ \Sigma_{\rm{x}}(t) \Sigma_{\rm{xx}}^{\dag}(t) 
- H. c. \right] U^{\dag} |0 \rangle
\nonumber\\
-\frac{1}{2}  \Sigma_{\rm{xx}}(t) \Sigma_{\rm{xx}}^{\dag}(t)
U^{\dag} |0 \rangle.
\label{matr}
\end{eqnarray}
In deriving Eq.\ (\ref{matr})
we used the fact that  
$\Sigma_{1}(t)$ annihilates the ground state $| 0 \rangle$ 
and that 

\begin{equation} 
\Sigma_{\rm{xx}}^{\dag}(t) | 0 \rangle=
\Sigma_{\rm{xx}}(t) U^{\dag}|0 \rangle=
\Sigma_{\rm{xx}}(t) \Sigma^{\dag}_{\rm{x}}(t) |0 \rangle=0 
\end{equation}
The latter relations can be obtained 
from Eqs.\ (\ref{sigin}) and (\ref{Uxx})
by noting that 
$H$ does not change the number of {\em e--h} pairs. 

It can be seen from 
Eqs.\ (\ref{matr})
that, since both 
$\Sigma_{\rm{xx}}^{\dag}(t)$ and $ \Sigma^{\dag}_{\rm{x}}(t)$ create a 
single {\em e--h} pair [see Eqs.\ (\ref{dir}), (\ref{Uxx}),  and
(\ref{sigin})],
the transition operator $U^{\dag}_{\rm{exc}}(t)$ 
also creates a single {\em e--h} pair.
Since, in addition,  the operators $\Sigma_{\rm{xx}}^{\dag}(t)$ and 
$ \Sigma^{\dag}_{\rm{x}}(t)$
conserve the momentum, as 
discussed above, we  have 

\begin{equation}
U^{\dag}_{\rm{exc}}(t)| 0 \rangle  
= \sum_{{\bf k}} M_{{\bf k}}(t) a^{\dag}_{{\bf k}} b^{\dag}_{-{\bf k}} | 0 \rangle.
\label{me2}
\end{equation}  
By comparing to the ``bare'' transition operator,
$U^{\dag}|0\rangle= \sum_{{\bf k}} a^{\dag}_{{\bf k}} b^{\dag}_{-{\bf k}}|0 \rangle$,
we see that the parameter

\begin{equation} 
M_{{\bf k}}(t) = \langle 0 | b_{-{\bf k}} a_{{\bf k}} U^{\dag}_{\rm{exc}}(t)|0\rangle,
\end{equation} 
should be interpreted as 
an effective time-dependent  optical transition matrix element. 

To derive an  explicit 
expression for
$M_{{\bf k}}(t)$, 
it is convenient to work in the exciton basis. 
We introduce the operators $B^{\dag}_{n,{\bf q}}$ and  
$B_{n,{\bf q}}$ that 
create/annihilate an exciton 
with center--of--mass momentum
${\bf q}$ and energy $E_{n{\bf q}}$ 
in the 
relative motion state $\phi_{n}({\bf q})$: 
 \begin{equation} 
B^{\dag}_{n,{\bf q}}
= \sum_{{\bf k}} \phi_{n}(k + \beta {\bf q}) 
\ a^{\dag}_{{\bf k}+{\bf q}} \, b^{\dag}_{-{\bf k}}, 
~~~~~~ \beta=\frac{m_{h}}{m_{e} + m_{h}}, \label{excit}
\end{equation}
with $m_{e}$ and $m_{h}$ being electron and hole mass, respectively,
and define the corresponding exciton contribution
to the linear polarization as
\begin{equation} 
P_{n}(t) = \sum_{{\bf k}} P_{\rm{x}}({\bf k},t) 
\phi_{n}^{*}({\bf k}).
\end{equation} 
The effects of the exciton--exciton interactions 
can be  conveniently described by 
introducing  
the correlation function

\begin{equation} 
\Sigma_{\rm{xx}}(n'm'{\bf q}';n;t)
= \frac{1}{2}\langle 0| B_{n'{\bf q}'} B_{m'-{\bf q}'}
\Sigma_{\rm{xx}}^{\dag}(t)
B^{\dag}_{n0}| 0 \rangle,
\label{sigme}
\end{equation} 
which represents 
the contribution to the  probability amplitude 
of pump--induced transitions between  
single--exciton,  
$B^{\dag}_{n0}| 0 \rangle$, and   two--exciton,
$B_{m'-{\bf q}'}^{\dag} B_{n'{\bf q}'}^{\dag} | 0 \rangle$,
states due to
 the  exciton--exciton interactions.
Such transitions 
also occur in the absence of interactions, 
with amplitude given  by 
Eq.\ (\ref{sigme}) but with $\Sigma_{\rm{xx}}^{\dag}(t)$ replaced by 
$\Sigma_{\rm{x}}^{\dag}(t)$.
This determines  the first (Phase Space Filling)
term in the rhs of Eq.\ (\ref{matr}). 
As we discuss below, by restricting in  Eq.\ (\ref{sigme})
to two--exciton states 
with zero center--of--mass momentum ${\bf q}'=0$, one recovers 
the static 
exciton--exciton interaction contribution
included in the SBE's.

Substituting 
$a^{\dag}_{{\bf k}} b^{\dag}_{-{\bf k}}
= \sum_{n} \phi_{n}^{*}({\bf k} ) 
B^{\dag}_{n,0}$ into 
 Eqs.\ (\ref{me2}) and (\ref{matr})
and inserting the closure relation 
into  the last term of Eq.\ (\ref{matr})
we obtain from Eq.\ (\ref{me2}) that 
$M_{{\bf k}}(t) = M_{{\bf k}}^{(\rm{PSF})}(t)
+  M_{{\bf k}}^{(\rm{xx})}(t)$, 
where 
$M_{{\bf k}}^{(\rm{PSF})}(t)
=  1 - \left|P_{\rm{x}}({\bf k},t)\right|^{2}$ 
is the usual Phase Space Filling 
and $M_{{\bf k}}^{(\rm{xx})}(t)$ comes  from 
the exciton--exciton interactions:

\begin{eqnarray}
 M_{{\bf k}}^{(\rm{xx})}(t)=
&&
\sum_{nn'm'} 
[ 
\Phi^{*}_{n'} P_{m'}(t) \phi_{n}({\bf k}) 
\Sigma_{\rm{xx}}^{*}(n'm'0;n;t)
\nonumber\\
&&
-\Phi^{*}_{n} P_{m'}^{*}(t) \phi_{n'}({\bf k}) 
\Sigma_{\rm{xx}}(n'm'0;n;t)
]
\nonumber\\
&&
-  \frac{1}{8} 
\sum_{mn'm'n{\bf q}'} \phi_{m}({\bf k}) 
\Phi^{*}_{n}
\Sigma_{\rm{xx}}^{*}(n'm'{\bf q}';m;t) 
\Sigma_{\rm{xx}}(n'm'{\bf q}';n;t),
\label{matrix}
\end{eqnarray} 
where $\Phi_{n} = \sum_{{\bf k}} \phi_{n}({\bf k})$.
The  first term in 
Eq. (\ref{matrix})
contains  both  static exciton--exciton interactions
(Hartree--Fock contribution) 
and exciton--exciton scattering processes
(correlation contribution). 
It describes a process where the photoexcitation of an {\em e--h} pair 
by the probe optical field is accompanied by the 
pump--induced creation and
subsequent annihilation of an {\em e--h} pair with zero center--of--mass
momentum.
Correlations contribute to this term via the  process during which  
the pump photoexcites an {\em e--h} pair with finite momentum
due to scattering with the probe--induced {\em e--h} pair, which  then 
scatters back to zero momentum 
and subsequently  annihilates. 
The last term in Eq.\ (\ref{matrix})
describes  exciton--exciton scattering processes (not
 included in the SBE's) 
where the pump photoexcites 
an {\em e--h} pair 
with finite  momentum 
due to scattering with the probe--induced {\em e--h} pair, which
subsequently annihilates by giving its momentum back to the
probe--induced {\em e--h} pair.

The contribution of the  biexciton transitions,
$| 0 \rangle \rightarrow U^{\dag}_{\rm{biexc}}(t) | 0 \rangle$,
to the pump--probe  polarization
is given  by Eq.\ (\ref{PP2}).
The biexciton optical transition operator 
$ U^{\dag}_{\rm{biexc}}(t)$, given by  Eq.\ (\ref{U2}),
can be obtained in a similar manner
by inserting the closure relation 
expressed in the basis of  momentum eigenstates
and noting that only two--pair states contribute. 
The result reads 
\begin{equation} 
U^{\dag}_{\rm{biexc}}(t) | 0 \rangle 
= \frac{1}{8} 
\sum_{n'm'q'n }  \Phi^{*}_{n}
\Sigma_{\rm{xx}}(n'm'{\bf q}';n;t) 
B^{\dag}_{n'{\bf q}'}  B^{\dag}_{m'-{\bf q}'} 
|0 \rangle,
\end{equation} 
where $\Sigma_{\rm{xx}}(n'm'{\bf q}';n;t)$ is given by 
Eq.\ (\ref{sigme}). Below we show that
$\Sigma_{\rm{xx}}(n'm'{\bf q}';n;t)$ 
can be expressed in terms of the two--exciton Green
function for the bare Hamiltonian $H$. 

\subsection{Effective Hamiltonian}

We now derive the second--quantized form of the effective Hamiltonian 
$H_{\rm{eff}}(t)$, whose formal expression is given by
Eq.\ (\ref{eff}). 
Here we primarily  focus on the excitonic contribution to 
the nonlinear polarization 
and therefore, 
according to Eq.\ (\ref{PP1}), we only need to 
consider how 
$H_{\rm{eff}}(t)$ acts  on single--pair states. 
 One has to distinguish between two contributions 
to the pump--induced correction
to the  Hamiltonian, 
$\frac{1}{2} \left([U, \Sigma_{1}^{\dag}] + H. c.\right)$, 
which come from
the decomposition  $\Sigma_{1}^{\dag}(t)
= \Sigma_{\rm{x}}^{\dag}(t) + \Sigma_{xx}^{\dag}(t)$. The first
contribution, due to 
the one--body single--exciton operator 
$\Sigma_{\rm{x}}^{\dag}(t)$, is also present 
in two--level systems.  Using Eq.\ (\ref{dir}), we obtain 
after straightforward algebra 

\begin{equation} 
\frac{1}{2} \left([U, \Sigma_{\rm{x}}^{\dag}] + H. c.\right)  
=\sum_{{\bf q}'}   {\rm Re} 
P_{\rm{x}}({\bf q}',t)
- \sum_{{\bf q}'} {\rm Re}  P_{\rm{x}}({\bf q}',t)
\left( 
 a^{\dag}_{{\bf q}'} a_{{\bf q}'} 
+ b^{\dag}_{-{\bf q}'} b_{-{\bf q}'}\right).\label{transop}
\end{equation} 
The second term in the rhs represents a correction to  the electron
and hole band energies (the first term 
plays no role and 
can be dropped). 
This  is  the origin of the  energy shifts 
(ac--Stark effect) 
well--known from  the dressed
atom picture.\cite{cohen,rev}  
Here, however, the shifts are momentum--dependent. This  leads, in
particular, to an additional  time-dependent correction to the 
exciton Bohr radius and wavefunction
absent within the two--level system approximation. 
\cite{ilias}  

To derive the many--body correction  to $H_{\rm{eff}}(t)$, described by 
$\Sigma_{\rm{xx}}^{\dag}(t)$, we note that, according to   
Eqs.\ (\ref{PP1}) and (\ref{me2}),
the time evolution operator ${\cal U}(t,t')$, and hence
$H_{\rm{eff}}(t)$, only act on the state
$U^{\dag}_{\rm{exc}}(t) | 0 \rangle$
that describes a single {\em e--h} pair with zero momentum.
Therefore,  it is sufficient 
to describe the action of $H_{\rm{eff}}(t)$  on the basis 
of states 
$a^{\dag}_{{\bf k}} b^{\dag}_{-{\bf k}} | 0 \rangle$. 
Since both $\Sigma_{\rm{xx}}^{\dag}(t)$ 
and $U^{\dag}$ conserve momentum
and 
create a single {\em e--h} pair,
we obtain that, when acting on states 
$ a^{\dag}_{{\bf k}}
b^{\dag}_{-{\bf k}}  | 0 \rangle$,

\begin{equation} 
U \Sigma^{\dag}_{\rm{xx}}(t)+ H.c.
=  \sum_{{\bf k}{\bf p}}
U_{\rm{eh}}({\bf p},{\bf k};t) a^{\dag}_{{\bf p}} 
b^{\dag}_{-{\bf p}}
b_{-{\bf k}} a_{{\bf k}}, 
\end{equation} 
where 

\begin{equation} 
U_{eh}({\bf p},{\bf k},t)=  
\langle 0| b_{-{\bf p}} a_{{\bf p}} 
\left[U \Sigma^{\dag}_{\rm{xx}}(t)+ H.c. \right] a^{\dag}_{{\bf k}}
b^{\dag}_{-{\bf k}}
 | 0 \rangle.
\end{equation} 
Following the steps described in the previous subsection
and transforming to the exciton basis, 
$U_{eh}({\bf p},{\bf k},t)$
can be presented as 

\begin{equation} 
U_{eh}({\bf p},{\bf k},t)= 
 \sum_{nn'm'}
\phi_{n'}({\bf p}) 
\phi_{n}^{*}({\bf k}) 
 \left[\Phi_{m'}  
\Sigma_{\rm{xx}}(n'm'0;n;t)
+ \Phi^{*}_{m'} 
\Sigma^{*}_{\rm{xx}}(nm'0;n';t)\right],
\label{pot}
\end{equation}
with $\Sigma_{\rm{xx}}(n'm'0;n;t)$ given by Eq.\ (\ref{sigme}).
Putting everything together,  we finally arrive at the following 
second--quantized expression for $H_{\rm{eff}}(t)$:

\begin{equation}
H_{\rm{eff}}(t) = 
\sum_{{\bf q}} \varepsilon_{{\bf q}}^{c}(t) a_{{\bf q}}^{\dag} a_{{\bf q}}
+  \sum_{{\bf q}} \varepsilon_{{\bf q}}^{v}(t) b_{-{\bf q}}^{\dag} 
b_{-{\bf q}} 
  -\sum_{{\bf k}k'{\bf q}} \,
\upsilon_{eh}({\bf k},{\bf k}';t)
a^{\dag}_{{\bf k}} \,
a_{{\bf k}'} \,  b^{\dag}_{-{\bf k}} \,  b_{-{\bf k}'},
\label{eff1} 
\end{equation}
where 

\begin{equation} 
\varepsilon_{{\bf q}}^{i}(t) = 
\varepsilon_{{\bf q}}^{i} + d_{p}(t) 
 {\rm Re}  P_{\rm{x}}({\bf q},t), ~~~ i=c,v,
\label{rband}
\end{equation} 
are the time-dependent  renormalized 
dispersion relations 
of  electrons  and holes, respectively, and
 
\begin{equation} 
\upsilon_{eh}({\bf k},{\bf k}';t)= 
\upsilon({\bf k}-{\bf k}') - d_{p}(t)U_{eh}({\bf k},{\bf k}';t)
\label{rint}
\end{equation} 
is the (nonlocal) effective {\em e--h} interaction.

It should be emphasized that the pump-induced parameter renormalizations in
$H_{\rm{eff}}(t)$  and $U^{\dag}_{\rm{exc}}(t)$ affect
the
polarization Eq.\ (\ref{PP1}) 
in different ways.
Indeed, the
dependence  of $H_{\rm{eff}}(t)$ on time follows that of the
pump pulse: 
the last terms  
in  Eqs.\ (\ref{rband}) and (\ref{rint})
are proportional to  $d_{p}(t)$. 
Therefore, 
they do not affect the 
time--evolution operator 
${\cal U}(t,t')$
when the probe comes at 
times  $t'$
{\em after} the pump pulse, but do contribute
to the pump--probe signal for negative
probe delays or when the probe overlaps in time with the pump
pulse. 
On the other hand, the transition operator Eq.\ (\ref{1})
depends on the pump only via $\Sigma^{\dag}_{1}(t)$ and hence, according to
Eq.\ (\ref{sigop}), is determined by 
the pump amplitude at all past times
$t' \le t$.
For positive time delays, the dynamics of the pump--probe spectra 
is therefore governed  by the renormalized transition 
matrix element  and the biexcitonic 
contribution Eq.\ (\ref{PP2}).  In this case, 
the latter is determined by the time--evolution operator 
corresponding to the bare Hamiltonian $H$.

We now discuss the role of the pump--induced 
corrections in the effective Hamiltonian 
$H_{\rm{eff}}(t)$.
As discussed in the Introduction, 
the spectral position and strength of the 
exciton resonance is determined by 
the time evolution 
operator ${\cal U}(t,t')$. 
In the linear response theory, this operator satisfies 
Eq.\ (\ref{evol}) with Hamiltonian  $H$, while in 
nonlinear absorption 
the latter must be  replaced by the effective Hamiltonian 
$H_{\rm{eff}}(t)$.
Therefore, 
$H_{\rm{eff}}(t)$ plays an important role in determining the 
properties of the excitonic resonance
observed in the pump--probe spectrum.  
First, the time--dependent change in the band dispersion relations  in
$H_{\rm{eff}}(t)$ leads to a shift in the position of the excitonic
resonance, which is, of course, the origin  of  the ac--Stark effect. 
Second, in the presence of
interactions, $H_{\rm{eff}}(t)$ does  not commute with itself 
at different times, which leads to memory effects. 
Importantly, such effects  only arise 
to second or higher order in the
Hamiltonian  correction $d_p(t)\Delta H(t)=
-\frac{1}{2}d_p(t)\left([U, \Sigma_{1}^{\dag}] + H. c.\right)$
and therefore  
do not contribute to the third--order polarizations. 
To illustrate this point, 
let us  expand ${\cal U}(t,t')$ in terms of the optical field 
using the Magnus expansion:\cite{muk1}

\begin{equation}\label{magnus}
{\cal U}(t,t') =
e^{-i H(t-t')/\hbar} 
\exp{\left[ \sum_{n=1}^{\infty} \frac{1}{n!} \left( -\frac{i}{\hbar}
\right)^{n} {\cal U}_{n}(t,t')\right]}, 
\end{equation} 
where 

\begin{eqnarray} 
{\cal U}_{1}(t,t') 
=
\int_{t'}^{t} 
dt_1 d_{p}(t_1) 
e^{i \frac{ H (t_1 - t') }{\hbar} } 
\Delta H(t_1) 
e^{-i \frac{ H (t_1 - t') }{\hbar} },
\end{eqnarray}

\begin{eqnarray} 
{\cal U}_{2}(t,t') 
&=&
\int_{t'}^{t} d_{p}(t_1) 
dt_1 \int_{t'}^{t_1}  dt_2
d_{p}(t_2)
\nonumber\\&\times&
\left[ e^{i \frac{ H (t_1 - t') }{\hbar} } 
\Delta H (t_1) e^{-i \frac{ H (t_1 - t') }{\hbar} } 
, e^{i \frac{ H (t_2 - t') }{\hbar} } 
\Delta H(t_2) e^{-i \frac{ H (t_2 - t') }{\hbar} }  \right], 
\label{evol1}
\end{eqnarray}  
etc. In the absence of Coulomb interactions, 
$[H,\Delta H(t)]=0$ and only the first term survives. 
${\cal U}_{1}(t,t')$ 
leads to the 
energy shifts if one does
not expand 
the exponential.
However, in order to describe the Coulomb--induced quantum 
dynamics 
emerging from the 
second and higher order terms in Eq.\ (\ref{magnus}), one cannot
restrict the pump--probe polarization to the third order in the optical
fields, but rather must 
use the full ${\cal U}(t,t')$.
Below we describe a simple physically--intuitive 
way, based on the similarity between 
$H_{\rm{eff}}(t)$ and $H$, for calculating the response to the probe
in all orders in the optically--induced terms
in $H_{\rm{eff}}(t)$. 

We start with the pump--probe polarization  Eq.\ (\ref{PP1}).
At time $t'$,
the probe photoexcites the system into a 
single--pair state 
$U^{\dag}_{\rm{exc}}(t')| 0 \rangle$, Eq.\ (\ref{me2}),
with the momentum wavefunction 
$M_{{\bf k}}(t')$.
Note that,
for negative time delays, $\Sigma_{1}(t')=0$ and hence 
$M_{{\bf k}}(t')=1$.
The subsequent evolution of such
state up to time $t$ is governed by $H_{\rm{eff}}(t)$ and 
describes the formation 
of the exciton resonance. 
The momentum wavefunction of the time--evolved state, $\Phi_{{\bf k}}(t)$,
satisfies the Schr\"{o}dinger equation

\begin{equation} 
i \hbar \frac{\partial \Phi_{{\bf k}}(t;t')}{\partial t}
= \left[ \varepsilon_{{\bf k}}^{c}(t) + \varepsilon_{{\bf k}}^{v}(t)
+ d_{p}(t) U_{\rm{eh}}({\bf k},{\bf k},t) \right] 
\Phi_{{\bf k}}(t;t')
- \sum_{{\bf k}'\ne {\bf k}}\upsilon_{eh}({\bf k},{\bf k}',t) 
\Phi_{{\bf k}'}(t;t'), \label{hyd} 
\end{equation} 
with the initial condition 
$\Phi_{{\bf k}}(t';t') =  M_{{\bf k}}(t')$.
Correspondingly, the polarization 
$ P_{{\bf k}_{\tau}}^{(\rm{exc})}(t)$, Eq.\ (\ref{PP1}),
takes the form 

\begin{equation} 
 P_{{\bf k}_{\tau}}^{(\rm{exc})}(t) =
\int_{-\infty}^{t} dt' d_{\tau}(t')
\sum_{{\bf k}} M_{{\bf k}}^{*}(t) 
\Phi_{{\bf k}}(t;t')  
\label{P1}
\end{equation}
Thus, polarization dephasing 
arises from  two different pump--induced effects:
the change in the 
transition matrix element
$ M_{{\bf k}}(t)$ 
and the time--evolution  
of the  state  
$\Phi_{{\bf k}}(t;t')$ 
due to  the time--dependent coefficients in the Schr\"{o}dinger 
equation Eq.\ (\ref{hyd}).
Note here that, for $E_{p}(t)=0$, 
$M_{{\bf k}}(t) = 1$ and Eq.\ (\ref{P1})
reduces to the familiar 
Fermi's Golden Rule expression. Note also 
that 
$M_{{\bf k}}(t)$ and the effective Hamiltonian 
parameters are time--independent in the case of monochromatic
photoexcitation.

The Hamiltonian $H_{\rm{eff}}(t)$ thus determines new (``dressed'')
exciton states, which depend on the time--dependent optical field
amplitude 
and, as in the linear response theory, can be obtained by solving the 
Schr\"{o}dinger equation Eq.\ (\ref{hyd}).
For example, this point of view provides a natural explanation 
of why, for below--resonant photoexcitation, 
the pump--induced coupling to higher exciton states
leads to a rigid exciton blue--shift 
instead of bleaching,
as would have been the case if only the $1s$ bound
state were included.\cite{ell,sbe,koch,knox2}
To see this, note that, to first approximation, 
the renormalization of the band dispersion relations,
Eq.\ (\ref{rband}), results in an increase
of  the e--h effective mass.
This in turn leads to an optically--induced decrease 
in the Bohr radius $a_{B}$, which, as is known from linear
absorption, directly determines the strength
of the exciton resonance. 
Since  the exciton strength is inversely proportional to  
$a_{B}^{d}$ ($d$ is the dimension), 
a decrease in Bohr radius
enhances the exciton strength and competes with the Phase Space
Filling reduction. Our approach also allows us to study the effect
of the pump on the exciton strength 
for short pulse durations,  where the adiabatic
limit does not apply, and compare it to the Fermi Edge
Singularity.\cite{ilias,knox2,acfes} This 
and other applications will be discussed elsewhere.

\subsection{Exciton--exciton interactions} 

In this section, we obtain the equation for the correlation function
$\Sigma_{\rm{xx}}(n'm'{\bf q}';n;t)$  which determines 
the renormalization of the parameters in the effective Hamiltonian
and the  transition 
matrix elements.

As can be seen from Eq.\ (\ref{sigme}), 
the time dependence of $\Sigma_{\rm{xx}}(n'm'q';n;t)$ is determined by
the operator $\Sigma^{\dag}_{\rm xx}(t)$.
In order to understand which 
processes contribute to $\Sigma_{\rm{xx}}(n'm'{\bf q}';n;t)$, it is
useful  to 
present $\Sigma^{\dag}_{\rm xx}(t)$, defined by Eq.\ (\ref{sigin}),  in 
second--quantized form:
 
\begin{equation} 
\Sigma_{\rm{xx}}^{\dag}(t)=
\sum_{(n,m) \ne (0,0)}
\frac{1}{(n+1)! \, n! \, (m+1)! \, m!}
\Sigma_{nm}^{\dag}(t), \label{op2}
\end{equation} 
where the operator 
\begin{eqnarray} 
\Sigma_{nm}^{\dag}(t)= 
\sum_{\{ {\bf p} {\bf  p}' {\bf  k}  {\bf k}' \}}
\Sigma_{nm}\left[\{{\bf p}';{\bf k}';{\bf k};{\bf p}\};t \right]
a^{\dag}_{{\bf p}'_{1}} \cdots a^{\dag}_{{\bf p}'_{n+1}}
b^{\dag}_{-{\bf k}'_{1}} \cdots b^{\dag}_{-{\bf k}'_{m+1}}
\nonumber\\
\times
b_{-{\bf k}_{m}} \cdots b_{-{\bf k}_{1}}
a_{{\bf p}_{n}} \cdots a_{{\bf p}_{1}} \label{op1}
\end{eqnarray} 
describes the photoexcitation of a single {\em e--h} pair accompanied by
its
 Coulomb interactions with 
$n$ additional photoexcited or Fermi sea 
electrons and $m$ additional 
photoexcited valence  holes. 

Substituting Eqs.\ (\ref{op1}) and (\ref{op2}) 
into Eq.\ (\ref{sig}), 
one can see that $\Sigma_{nm}^{\dag}(t)$ couples only to 
$\Sigma_{n-1m}^{\dag}(t)$ and $\Sigma_{nm-1}^{\dag}(t)$ 
but not to terms with higher $n$ or $m$. 
This defines an iterative scheme
for obtaining successively higher 
amplitudes. 
As a result, 
in  undoped semiconductors, 
only a {\em finite} number of low--order  terms 
$\Sigma_{nm}^{\dag}(t)$ contribute to 
the nonlinear optical response to a given order 
in the optical fields. 
For example, 
only the following 
interaction terms 
contribute 
to the third--order nonlinear polarizations: 
$\Sigma_{10}^{\dag}(t)$ 
(describing interactions between the photoexcited {\em e--h} pair and an 
additional electron), 
$\Sigma_{01}^{\dag}(t)$ 
(describing interactions between the photoexcited {\em e--h} pair and an 
additional hole), 
and $\Sigma_{11}^{\dag}(t)$ 
(describing interactions between the photoexcited {\em e--h} pair and an 
additional electron and hole). 
Thus, a calculation of the third--order nonlinear 
polarization requires the solution of a four--body problem.  
Higher--order interaction terms 
do not contribute 
because the corresponding $\Sigma_{nm}^{\dag}(t)$ give zero
when acting on 
states 
with  up to two {\em e--h} pairs.

We now derive  the equation
for $\Sigma_{\rm{xx}}(n'm'{\bf q}';n;t)$, which 
is more convenient for describing the above processes in the case of
excitons. 
Starting  from the definition 
Eq.\ (\ref{sigme}), using 
Eq.\ (\ref{sigin})
for $\Sigma_{\rm{xx}}^{\dag}(t)$, 
and going to the basis  of exciton eigenstates of
$H$, we obtain 
after some algebra

\begin{eqnarray}
\Sigma_{\rm{xx}}(n'm'{\bf q}';n;t)= 
&&
-\frac{i}{\hbar}\int_{-\infty}^{t} dt' 
e^{- i\left(\Delta \Omega - E_{n0} \right)(t-t') / \hbar}
\nonumber\\ && \times
\sum_{n''m''{\bf q}''}
 G(m'n'{\bf q}';m''n''{\bf q}'';t-t') 
\upsilon({\bf q}'') \ F_{nn''}({\bf q}'')
\nonumber\\ && \times
\sum_{{\bf k}'} \left[ P_{\rm{x}}({\bf k}',t')  - P_{\rm{x}}({\bf
k}'-
{\bf q}'',t')\right] 
\phi_{m''}^{*}({\bf k}' - \beta {\bf q}'')
\label{sigme2}
 \end{eqnarray} 
where 

\begin{equation} 
F_{nm}({\bf q}) = \sum_{{\bf k}} \left[ \phi_{n}({\bf k}+{\bf q}) 
 - \phi_{n}({\bf k}) \right] 
\phi_{m}^{*}({\bf k} +\beta {\bf q})
\end{equation} and 
$G(m'n'{\bf q}';m''n''{\bf q}'';t)$
is the two--exciton Green function\cite{muk1,sham}  
 defined as 
\begin{equation} 
 G(m'n'{\bf q}';m''n''{\bf q}'';t) 
= \langle 0 | 
B_{n',{\bf q}'} B_{m',-{\bf q}'} 
B^{\dag}_{m'',-{\bf q}''}(t)  B^{\dag}_{n'',{\bf q}''}(t) |0 \rangle,
\label{green}  
\end{equation}
with  $ B^{\dag}_{m,{\bf q}}(t)= 
e^{-i H t } B^{\dag}_{m,{\bf q}} e^{i H t /\hbar}$.
Thus, the problem of calculating the effective parameters 
is reduced to 
the two--exciton Green function Eq.\ (\ref{green}) with the ``bare''
Hamiltonian $H$.
The Dyson equation for $G$ can be obtained 
in a straightforward way starting from the equations of motion 
for the exciton Heisenberg operators 

\begin{eqnarray}
 G(m'n'{\bf q}';mn{\bf q};t) 
&&
=e^{-\frac{i}{\hbar}
 \left( E_{n,-{\bf q}} + 
E_{m,{\bf q}}\right) t} 
\langle 0 | 
B_{n',{\bf q}'} B_{m',-{\bf q}'} 
B^{\dag}_{m,-{\bf q}} B^{\dag}_{n,{\bf q}}
 | 0 \rangle  
\nonumber\\
&&
+\frac{1}{i \hbar} 
\int_{0}^{t}  dt' 
e^{\frac{i}{\hbar} \left( E_{n,-{\bf q}} + E_{m,{\bf q}}\right) (t'-t)} 
\nonumber\\ &&
\times\sum_{m'',n'',{\bf q}''} V_{n'' m''}^{nm}({\bf q}''-{\bf q})
G(m'n'{\bf q}';m''n''{\bf q}'';t'), 
\label{eom}
\end{eqnarray} 
where 
$ V_{n'm'}^{nm}({\bf q})
= \upsilon_{{\bf q}} 
F_{nn'}({\bf q}) F_{mm'}(-{\bf q})$
is the exciton--exciton interaction potential.

The  Hartree--Fock approximation corresponds to 
retaining 
only  the first term in the rhs 
of Eq.\ (\ref{eom}). 
Substituting the latter into Eq.\ (\ref{sigme2})
and using our previous results for the effective transition 
matrix element and Hamiltonian parameters, 
one obtains  
the contribution to the nonlinear polarization 
that is of   first order in the exciton--exciton interaction. 
It can be seen that, in this case,  
only two--exciton states with zero center--of--mass momentum 
contribute to the nonlinear polarizations.  
The correlation effects due to exciton--exciton scattering 
involving the exchange of center--of --mass momentum are 
described by the second term 
in the rhs 
of Eq.\ (\ref{eom}). 
To  lowest order 
in the exciton--exciton interaction, these 
can be obtained by solving 
Eq.\ (\ref{eom}) iteratively.
This is equivalent to a second-- or higher--order 
Born approximation to the self--energy 
that determines the effects of the exciton--exciton interactions 
on the nonlinear optical polarization. 
In the case where opposite--spin excitons are photoexcited, 
Eq.\ (\ref{eom}) can also be used to study the effects  of bound biexciton 
states.

As can be seen from 
Eq.\ (\ref{sigme2}), 
the two--exciton Green function acts as a memory kernel 
that  determines the 
effective parameters at time $t$
in terms of the linear polarization at earlier times.  
If one only retains  the first term on the rhs of Eq.\ (\ref{eom})
(Hartree--Fock approximation), 
such memory 
effects decay exponentially 
as determined by the dephasing time. 
The significance  of the correlation term  
in Eq.\ (\ref{eom}) is that, under certain conditions, 
the memory kernel can become 
non--exponential (non--Markovian memory effects). 
The importance  of such effects 
for short pulse excitation 
will be discussed  elsewhere. 
Note here that in the case of monochromatic excitation, for which 
$\Sigma_{\rm{xx}}^{\dag}$ is time--independent, 
the correlations 
simply change   the magnitude 
of the various parameters.

\section{Four--Wave--Mixing  Polarization} 

Let us now turn to the polarization along the four--wave--mixing
direction, ${\bf k}_{3} = 2 {\bf k}_{p} - {\bf k}_{\tau}$.
The  latter can be obtained by substituting Eqs.\ (\ref{UU1}) and
(\ref{UU2}) into  Eq.\ (\ref{pol}) 
and retaining only terms proportional to 
$e^{i {\bf k}_{3} {\bf r}}$.
After some algebra, we obtain 

\begin{eqnarray} 
P_{\rm{FWM}}(t) =  \frac{i\mu}{ \hbar}  
e^{i {\bf k_{3}}  {\bf r} - \omega_{p} (t-\tau)}
\int_{- \infty}^{t} d_{\tau}(t')  
 \langle 0| [U {\cal U}(t',t) U^{\dag}_{\rm{FWM}}(t)-
(t\leftrightarrow t')] 
|0 \rangle, \label{fwm1} 
\end{eqnarray} 
where 
\begin{equation} 
U^{\dag}_{\rm{FWM}}(t) 
= U W^{\dag}_{\rm{pp}}(t) - \Sigma_{\rm{x}}^{\dag}(t) U
\Sigma_{\rm{x}}^{\dag}(t), 
\end{equation}
and 

\begin{equation}
W_{\rm{pp}}^{\dag}(t) =
\frac{1}{2} \left[\Sigma_{1}^{\dag}(t)\right]^{2} - 
\Sigma_{2}^{\dag}(t),
\label{irreduce}
\end{equation} 
is the ``irreducible'' biexciton operator satisfying 

\begin{equation}
i \hbar \frac{\partial }{\partial t}W^{\dag}_{\rm{pp}}(t)|0 \rangle= 
H     W_{\rm{pp}}^{\dag}(t) | 0 \rangle - \frac{d_{p}(t)}{2}
U^{\dag} \Sigma_{\rm{x}}^{\dag}(t) | 0 \rangle.
\label{biexc} 
\end{equation}
In the above, we used the fact the $\Sigma^{\dag}_{\rm{xx}}(t) 
| 0 \rangle =0$.
Thus, the four wave mixing polarization is determined by 
$\Sigma_{\rm{x}}(t)$, which describes 
the linear pump--induced polarization, 
and the biexciton 
transition operator $W_{\rm{pp}}(t)$, which includes all 
exciton--exciton interaction effects 
in the four--wave--mixing signal. 
One can express 
$W_{\rm{pp}}^{\dag}(t)| 0 \rangle$ in terms of the two--exciton Green
function Eq.\ (\ref{green})
by following the procedure of the previous section.
Note here that, unlike for the pump---probe polarization,  
the pump--induced corrections to the 
Hamiltonian  do not contribute to the 
third--order polarization
along the direction ${\bf k_{3}}$.
Therefore, the $\chi^{(3)}$ 
calculations of the four--wave--mixing signal 
neglect the optically--induced change in the 
time evolution 
operator ${\cal U}(t,t')$, 
while similar pump--probe 
calculations 
include this effect only to first order  in perturbation theory.
As discussed in the previous section, such calculations do not
properly treat  the  
shift of the exciton resonance
and neglect the memory effects and quantum dynamics 
induced by the interactions in $H_{\rm{eff}}(t)$. 
The latter are especially important in the case of the 
Fermi--edge singularity.\cite{ilias}

Finally, let us discuss the connection between our approach and other
works.
In Refs. 13 and 14, 
it was shown that the third--order nonlinear polarization can be
expressed exclusively 
in terms of single--pair and two--pair creation operators,
corresponding to our  $\Sigma^{\dag}_{\rm{x}}(t)$ 
and $W^{\dag}_{\rm{pp}}(t)$.
Similarly, one can obtain  $\chi^{(3)}$ along the 
four--wave--mixing direction 
by simply replacing  $H_{\rm{eff}}(t)$ in  Eq.\ (\ref{fwm1})
by $H$.
The connection to  Ref. 14 
can be established by noting that 
$\langle \alpha 2|W_{\rm{pp}}^{\dag}(t)|0 \rangle $ 
coincides with  the lowest term in the expansion 
of the Hubbard 
operator $|0 \rangle \langle \alpha 2|$.
Furthermore, by projecting  Eq.\ (\ref{biexc}) into two-pair states, 
one can identify the amplitude 
$\langle  0| b_{-{\bf k}} b_{-{\bf k}'} a_{{\bf k}'-{\bf q}} 
a_{{\bf k}+{\bf q}}
W_{\rm{pp}}^{\dag}(t) | 0 \rangle$ 
with the four--point density matrix 
introduced by Axt and Stahl.\cite{eom} 
This connection is less obvious 
for our pump--probe results, 
Eqs. \ (\ref{PP1}) and (\ref{PP2}). 
The reason is that, as discussed above, 
these include important contributions 
beyond $\chi^{(3)}$. 
However, by only keeping the lowest term in a perturbative 
expansion of 
the time--evolution operator 
[i.e., by restricting to $\chi^{(3)}$], 
one  can obtain
after  straightforward but laborious algebra 
the expression for the pump--probe polarization exclusively 
in terms of 

\begin{equation} 
\Sigma_{\rm{j}}^{\dag}(t)|0\rangle 
= \frac{i}{\hbar} 
\int_{-\infty}^{t} dt' d_{j}(t') 
e^{-i H (t-t')/\hbar}
U^{\dag}|0 \rangle,
\label{sig11}
\end{equation}
and 

\begin{equation}
W_{\rm{ij}}^{\dag}(t) | 0 \rangle=
\frac{i}{\hbar}\int_{-\infty}^{t } dt'd_{i}(t') 
e^{- i H(t-t')/\hbar} \
U^{\dag} \Sigma^{\dag}_{\rm{j}}(t') |0 \rangle,
\label{w11}
\end{equation}
with i,j=$p,\tau$. The corresponding expression is rather tedious
and is  omitted here.
By comparing Eq.\ (\ref{sig11})  to 
Eqs.\ (\ref{sigop}) and\ (\ref{dir}),
it can be seen that 
$\Sigma_{\rm{p}}^{\dag}(t)|0 \rangle 
= \Sigma_{\rm{x}}^{\dag}(t) |0 \rangle$. 
It can also be seen from  Eq.\ (\ref{w11})
that  $W_{\rm{pp}}^{\dag}(t) | 0 \rangle$ 
coincides with  
the solution of 
Eq.\ (\ref{biexc}). 
The third--order  polarization 
is thus determined by 
the states $W_{\rm{ij}}^{\dag}(t) | 0 \rangle$, which, 
for  $i \ne j$,
describe the effects 
of  the interference
between the pump and the probe.

\section{ CONCLUSIONS}

In conclusion, we discussed a general 
theoretical approach for describing Coulomb many--body effects in
the ultrafast 
coherent dynamics in  semiconductors and metals. 
Our approach provides a complete description of the correlation 
and non--Markovian memory effects by reducing
the  calculation of the 
nonlinear optical response
of the ``bare'' system
onto  the linear response 
of a ``dressed'' 
system described by the effective time-dependent Hamiltonian,
Eq.\ (\ref{eff1}).
Furthermore, it allows us to study 
the coherent ultrafast nonlinear optical 
response of Fermi sea systems.\cite{ilias} 

As is known,\cite{muk} the calculations 
of the third--order nonlinear polarizations 
starting from equations of motion 
must address the time evolution of exciton and biexciton states. 
In view of the 
unresolved numerical problems facing 
a straightforward calculation of 
biexcitonic effects (except in one dimension), 
our approach has the advantage 
of providing  
a physically-intuitive picture 
for viewing their manifestations 
in the  ultrafast nonlinear optical response. 
Close to the exciton frequency, 
these effects can be taken into account 
via the renormalization on the 
electron/hole dispersions and 
interaction potential as well as the probe--induced 
optical transition matrix elements. 
Our approach also allows us to treat the 
Coulomb--induced quantum dynamics arising from the 
semiconductor response to such perturbations, 
which cannot be captured within the 
$\chi^{(3)}$ approximation. 
Note that, for monochromatic excitation, 
the exciton--exciton correlations 
simply change the magnitude of the effective parameters, 
while for ultrashort pulses they also affect  their  
time--dependence over times comparable 
to the dephasing time. 		

Our theory separates 
the effects due to the renormalization  of the 
electron/hole dispersions and 
interaction potential 
from the renormalization of 
the probe--induced 
optical transition matrix elements. 
These affect the nonlinear polarization 
in a different way, as can  already be seen in the absence of Coulomb 
interactions by  recalling the 
well--known ``dressed atom'' results 
for a two--level system.\cite{cohen}  In the latter case, 
the shift in the energies results in an overall resonance shift
(e.g., ac--Stark effect) 
which cannot be correctly described within  $\chi^{(3)}$
calculations.
Such effects are described here 
by the time evolution operator for the 
effective Hamiltonian 
$H_{\rm{eff}}(t)$. 
In semiconductors, 
the additional
quantum dynamics (not captured by $\chi^{(3)}$)
arising from  the non--commutativity of the effective Hamiltonian
at different times
can be simply described  
with the 
Schr\"{o}dinger equation for 
the effective exciton Hamiltonian 
$H_{\rm{eff}}(t)$. 
On the other hand, the renormalization 
of the transition matrix elements,
due to Phase Space Filling and, in semiconductors, 
exciton--exciton interaction effects,
leads to an overall factor  in 
the polarization that can be calculated to 
third--order in the optical fields.

Thus our theory identifies a number of 
competing  physical processes, 
each leading to different nonlinear dynamics and spectral 
features. 
In undoped semiconductors, 
we can obtain the parameters determining the strength of the 
above nonlinearities in terms of  the ``bare'' 
semiconductor one-- and two--exciton 
Green functions.
Finally, we also established a
connection between 
our approach and $\chi^{(3)}$ 
calculations in undoped semiconductors.
We believe that the present 
theory is well-suited for 
interpreting the physics conveyed by ultrafast  coherent 
nonlinear optics experiments
in semiconductors and metals 
using pulses as short as 10 fs.

\acknowledgements

This work was supported by NSF CAREER award ECS-9703453, and, in
part, by ONR Grant N00014-96-1-1042  and by HARL, Hitachi Ltd.

\end{document}